\documentclass[journal,
                    preprint
                    ]{revtex4}

\usepackage{epsfig}
\usepackage{color}

\begin{document}

\def\e{\begin{equation}}
\def\f{\end{equation}}
\def\=#1{\overline{\overline #1}}
\def\s{\strut\displaystyle}
\def\_#1{{\bf #1}}
\def\o{\omega}
\def\v{\varepsilon}
\def\M{\mu}
\def\M0{\mu_0}
\def\D{\nabla}
\def\.{\cdot}
\def\x{\times}
\def\##1{{\bf#1\mit}}
\def\Re{{\rm Re\mit}}
\def\Im{{\rm Im\mit}}
\def\l#1{\label{eq:#1}}
\def\r#1{(\ref{eq:#1})}
\def\c#1{\cite{#1}}
\def\vec#1{{\bf #1}}

\parskip=0pt
\parindent=9mm

\abovecaptionskip=-0.5pt

\title{Impedance-matched microwave lens}

\author{Pekka Alitalo, Olli Luukkonen, Joni Vehmas,
        Sergei~A.~Tretyakov\\
Department of Radio Science and Engineering\\SMARAD Center of
Excellence\\ TKK Helsinki University of Technology\\ P.O. Box
3000, FI-02015 TKK, Finland.\\ e-mail: pekka.alitalo@tkk.fi
\vspace{20pt}}

\begin{abstract}
A microwave lens with highly reduced reflectance, as compared to
conventional dielectric lenses, is proposed. The lens is based on
two-dimensional or three-dimensional transmission-line networks
that can be designed to have an effective refractive index larger
than one, while having almost perfect impedance matching with free
space. The design principles are presented and an example lens is
studied using commercial simulation software.
\end{abstract}


\maketitle

\section{Introduction}

Homogeneous dielectric materials having refractive index $n$
different from that in free space ($n=1$ in free space), have been
used for a long time in microwave lens
applications~\cite{Antenna_Theory,Microwave_Optics,Lo}. The
benefits of such lenses are their simple structure and
straightforward design. One of the main drawbacks of such lenses
is the evident impedance mismatch with the material surrounding
the lens. This causes reflections from the lens and, thus,
unwanted loss of power.

In this letter, we propose a novel method to overcome this
drawback. The lens design which is introduced here is based on the
use of transmission-line (TL) networks that have a different
propagation constant than the surrounding medium, e.g., free
space. Because the network impedance can be designed separately
from the refractive index (unlike in a dielectric lens), the
impedance matching with free space or some other medium can be
obtained for different refractive indices by properly choosing the
filling material and geometry of the transmission lines comprising
the network. Other well-known methods to reduce reflections from a
dielectric lens are e.g. restricting the refractive index to very
low values, introduction of matching layers (analogous to
quarter-wave transformers), and using inhomogeneous
materials~\cite{Antenna_Theory}. As compared to the method
proposed here, the previous methods suffer either from very narrow
operation bandwidth or complicated and expensive manufacturing.

Lens structures based on the use of transmission lines or
waveguides have been widely reported in the literature, see,
e.g.,~\cite{Lo}. For example, structures consisting of stacked
parallel metal plates, or arrays of parallel metal wires, behaving
as artificial dielectrics with the permittivity less than one,
have been used to create lenses. These lenses suffer from the same
impedance matching problem as the dielectric ones. In addition,
their effective refractive index strongly depends on the
frequency. Other examples of previously proposed lenses (based on
the use of transmission lines) use arrays of transmission lines in
order to create the needed phase shift enabling the correct
refraction angle at the lens surface~\cite{Rotman,Patton}. The
difference with the lens studied in this letter is that the
interfaces of the lenses discussed in~\cite{Rotman,Patton} behave
effectively as antenna arrays and waves inside these lenses
propagate only along one direction, whereas the lens discussed in
this letter can be studied and modelled in a way similar to a
homogeneous dielectric lens, since the period of the structure is
much smaller than the wavelength and wave propagation is
effectively isotropic.

Furthermore, the introduction of the transmission-line approach to
the lens design enables easy manufacturing of inhomogeneous lenses
and even electrically controllable lenses, since the TL networks
can be loaded by lumped elements (e.g. capacitors and/or
inductors) and controllable elements such as varactors. In this
letter we concentrate on a simple, i.e., an unloaded
two-dimensional network, in order to demonstrate the feasibility
of the proposed method.

The principle of the proposed approach to microwave lens design
has been recently presented~\cite{Alitaloiwat08}. In the current
letter we show in detail how a lens comprised of stacked networks
of transmission lines can be designed in the same way as a
traditional dielectric lens. We compare an example
transmission-line lens to a reference dielectric lens and
demonstrate the improvement of the impedance matching by using
commercial simulation software. Also the focusing characteristics
of the lenses are compared, validating the use of the lens design
equations.

\section{Lens design}

The impedance mismatch related to traditional dielectric lenses
occurs because the refractive index $n=\sqrt{\varepsilon_r \mu_r}$
is larger than unity due to the relative permittivity
$\varepsilon_r>1$, while the relative permeability $\mu_r$ is
equal to unity as in free space. Therefore the wave impedance of
the dielectric material differs from that in free space by the
factor $\sqrt{\frac{\mu_r}{\varepsilon_r}}$. One solution to the
impedance mismatch problem would be the use of magneto-dielectric
materials having $\mu_r>1$. However, in the microwave and
millimeter-wave range the use of magneto-dielectrics is either
impossible or prohibitively expensive.

As shown e.g. in~\cite{Eleftheriades_book,Caloz_book,AlitaloJAP},
a two-dimensional or a three-dimensional transmission-line network
can be treated as an effective medium with a certain effective
wavenumber for waves travelling inside the network. As long as the
period of the network is significantly smaller than the wavelength
of waves in the network, the network is effectively
isotropic~\cite{Alitalo_metmat}. The dispersion in an unloaded
network can be derived from the rules for voltages and currents
travelling in the
network~\cite{Eleftheriades_book,Caloz_book,AlitaloJAP}.
In~\cite{AlitaloJAP} the dispersion and impedance were studied for
three-dimensional unloaded and loaded networks, but in this letter
we choose to use an unloaded two-dimensional network, for
simplicity of consideration and design. The dispersion relation of
such a network is~\cite{AlitaloTAP}

\e \cos(k_{\rm x}d)+\cos(k_{\rm y}d)=4\cos^{2}({k_{\rm TL}d/2)}-2,
\l{disp} \f where $k_{\rm x,y}$ is the wavenumber in the network
along axes ($\rm x,y$), $d$ is the period, and $k_{\rm TL}$ is the
wavenumber of waves in the transmission lines. The wavenumber in
the network is $k=\sqrt{k_x^2+k_y^2}$.

\begin{figure}[t!]
\centering {\epsfig{file=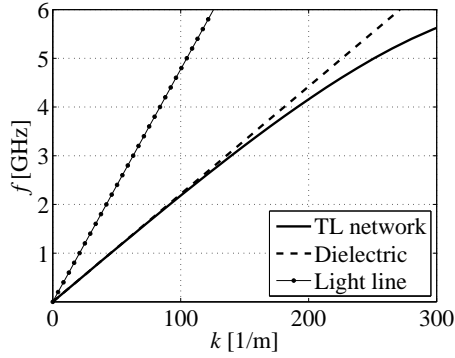,width=0.4\textwidth}}
\caption{Dispersion in a transmission-line (TL) network for axial
propagation ($k_y=0$, $k=k_x$), in a homogeneous dielectric
material with $\varepsilon_r=4.66$ and in free space (``light
line'').} \label{dispersion}
\end{figure}

Using~\r{disp} we can plot the dispersion curve of a network and
find the value of the relative permittivity of a reference
dielectric having the same dispersion in the desired frequency
band. In Fig.~\ref{dispersion} we plot the dispersion diagram for
an example network with $d$=8~mm and $k_{\rm TL}=c_0/\sqrt{2.33}$
($c_0$ is the speed of light in vacuum, and we assume that the
dielectric material filling the transmission lines of the network
has the relative permittivity equal to $\varepsilon_{r,\rm
TL}=2.33$). Fig.~\ref{dispersion} also shows the dispersion curves
for a wave in free space (the light line) and in a homogeneous
dielectric material with $\varepsilon_r=2\varepsilon_{r,\rm
TL}=4.66$. Comparing the plots, we can conclude that at lower
frequencies, i.e., below 3~GHz in this case, the refractive index
of the TL network, which is defined by the wavenumber, is close to
that of a dielectric material with the refractive index
$n=\sqrt{\varepsilon_r}=\sqrt{4.66}$. The studied network is also
effectively isotropic approximately below 3~GHz, since for the
diagonal propagation ($k_x=k_y$), the dispersion curve is the same
as for the dielectric material with $\varepsilon_r=4.66$ (dashed
line in Fig.~\ref{dispersion}).

The impedance of the studied network can be derived from the
equations for a three-dimensional unloaded network, presented e.g.
in~\cite{AlitaloJAP}. The network impedance in the two-dimensional
case reads

\e Z=\frac{jZ_{\rm TL}\sin({k_{\rm TL}d/2)
(1+e^{-jkd})}}{\cos({k_{\rm TL}d/2) (1-e^{-jkd})}}=Z_{\rm
TL}\frac{\tan(k_{\rm TL}d/2)}{\tan(kd/2)}, \l{imped} \f where
$Z_{\rm TL}$ is the impedance of isolated sections of transmission
line. Using~\r{imped} we have found that in order to obtain the
network impedance of $120\pi$~$\Omega \approx 377$~$\Omega$ (the
free-space impedance) approximately at 3~GHz, the impedance of the
TL sections comprising the network should be designed to have
impedance of $Z_{\rm TL}=585$~$\Omega$. The resulting network
impedance as a function of the frequency, plotted using~\r{imped},
is shown in Fig.~2. The impedance curve demonstrates that the
network impedance varies quite smoothly and, therefore, good
impedance matching with free space is expected in a relatively
large frequency band around 3~GHz.\newline

\begin{figure}[t!]
\centering {\epsfig{file=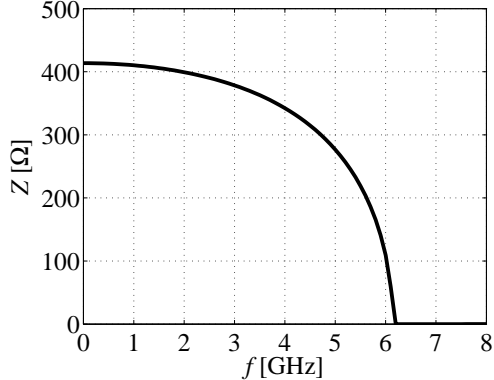,width=0.4\textwidth}}
\caption{Impedance of the studied example transmission-line
network as a function of frequency.} \label{impedance}
\end{figure}

\section{Full-wave simulations of a realizable lens}

As discussed above, the example TL network has the period of
$d=8$~mm, $k_{\rm TL}=c_0/\sqrt{2.33}$, and the impedance (of the
TLs) of 585~$\Omega$. To obtain this, we have decided to use TLs
made of parallel metal strips, embedded in a background material
with $\varepsilon_{r,\rm TL}=2.33$. Using the simple
parallel-plate approximation, as was done
in~\cite{Alitaloiwat08,AlitaloTAP}, we have found that the
suitable width and separation of the TLs are 1.266~mm and 3~mm,
respectively~\cite{Alitaloiwat08}.

The designed lens structure has been simulated with Ansoft High
Frequency Structure Simulator (HFSS) software. First, to make sure
that good impedance matching is obtained, the network is simulated
as a transversally infinite (infinitely periodic in $y$- and
$z$-directions) slab with a normally incident electromagnetic
plane wave, having electric field parallel to the $z$-axis,
illuminating the slab. The simulation model, with the thickness of
the TL network equal to $8d$, is shown in Fig.~3.

\begin{figure}[b!]
\centering {\epsfig{file=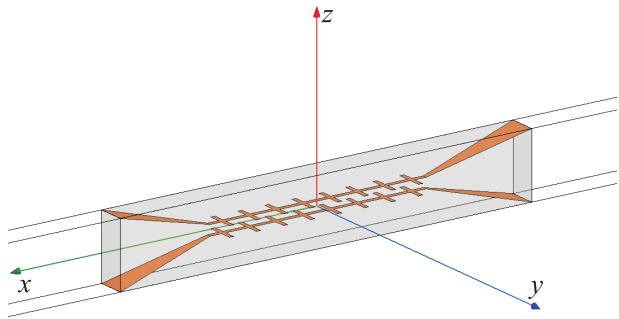, width=0.5\textwidth}}
\caption{Color Online. HFSS simulation model of the transversally
infinite slab. The edges of the simulation model are assigned as
perfect electric conductors ($xy$-planes) and perfect magnetic
conductors ($xz$-planes) to obtain periodicity. The metal strips
are modelled as infinitely thin and perfectly conducting.}
\label{1dmodel}
\end{figure}

To couple waves from free space to the network and vice versa, we
have introduced short sections of gradually enlarging transmission
lines, as proposed in~\cite{AlitaloTAP}. These lines of the
``transition layer'' have the length of 30~mm and the ratio
between the width and height equal to the TLs in the network. At
the end of these lines (at the interface with free space), the
width (along the $y$-axis) and separation (along the $z$-axis) of
the lines are 8.0~mm and 18.96~mm, respectively.

\begin{figure}[t!]
\centering {\epsfig{file=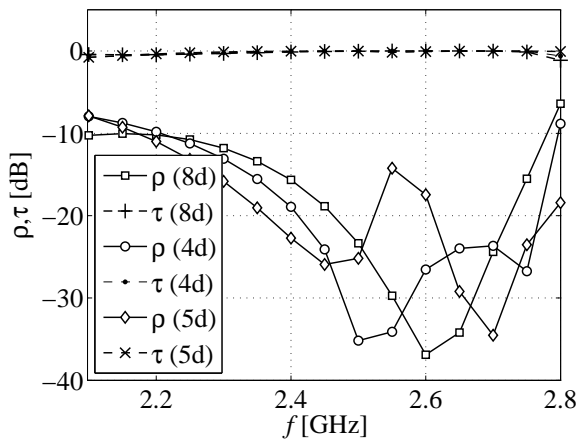, width=0.5\textwidth}}
\caption{Simulated reflection ($\rho$) and transmission ($\tau$)
for the slab shown in Fig.~3 (thickness of the slab being $8d$),
as well as for two other slabs having different thicknesses.}
\label{rho_tau_1d}
\end{figure}

The simulation results for the transversally infinite slab, shown
in Fig.~4, demonstrate good impedance matching with free space. To
illustrate the fact that changing the slab thickness does not
destroy the impedance matching, we show results also for slabs
with the thickness of the TL network being $4d$ and $5d$. Although
the TL network impedance should be optimally matched with free
space around 3~GHz, we have found the optimal frequency to be
around 2.5~GHz, as seen in Fig.~\ref{rho_tau_1d}.
In~\cite{Alitaloiwat08} for the same network, the optimal
impedance matching was obtained at 3.5~GHz. The reason for the
difference in these two cases is that the impedance of the
transition layer was designed to be around 377~$\Omega$
in~\cite{Alitaloiwat08}, whereas in this letter we have used the
same impedance as in the TLs comprising the network
($\sim$585~$\Omega$). Although the transition layer is short
compared to the wavelength, the specific design of the transition
layer clearly affects the frequency dependence of the impedance
matching and therefore it should be taken into account in the
design of a lens for specific application, e.g., by optimization
using a suitable simulation software.

\begin{figure}[t!]
\centering {\epsfig{file=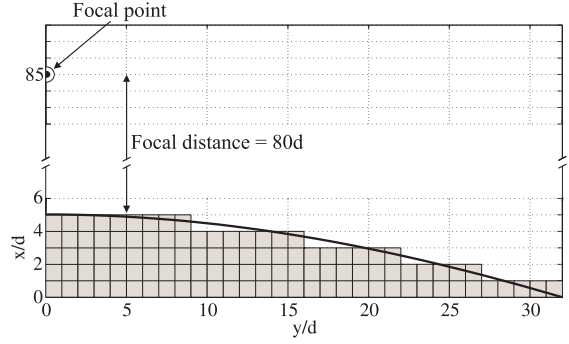,width=0.45\textwidth}}
\caption{Analytically calculated lens curvature (solid black line)
and the transmission-line network (each square represents one unit
cell of the network) having approximately the same curvature. Only
half of the lens is shown along the $y$-axis since the simulation
model is cut in half by a PMC boundary.} \label{lens}
\end{figure}

To demonstrate the functionality of the proposed lens, a model
illustrated in Fig.~\ref{lens} was simulated by illuminating the
lens with a normally incident plane wave. Here we use such
boundary conditions that we effectively simulate a structure which
is infinitely periodic in the vertical direction, i.e., along the
$z$-axis. Also, a perfect magnetic conductor (PMC) boundary is
assigned to the $xz$-plane in the center of the lens to reduce the
simulation time. The lens curvature is designed using the
well-known lens design equations for dielectric lenses~\cite{Lo}
to obtain a lens that focuses a plane wave to a line on the other
side of the lens. One side of the lens (the side from where the
plane wave impinges on the lens) is flat and the other side has a
certain curvature. Here we have assumed that the lens material has
a refractive index of $n=\sqrt{4.66}$. The width of the lens along
the $y$-axis is $64d=512$~mm and with this width and curvature,
the focal point is at a distance of $80d$ from the lens~\cite{Lo}.
For the practical implementation of the TL network, the optimal
curvature is approximated by a stepwise structure as shown in
Fig.~\ref{lens}.

\begin{figure} [b!]
\centering {\epsfig{file=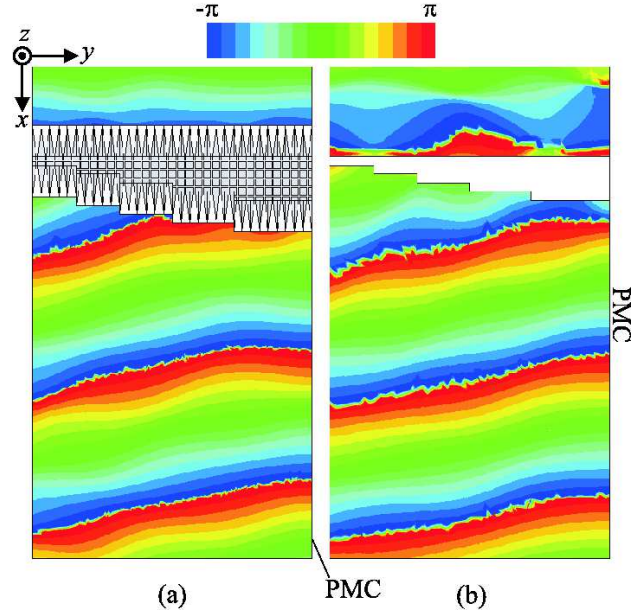, width=0.5\textwidth}}
\caption{Color Online. Phase of the simulated electric field (at
the frequency 2.4~GHz) for (a)~the transmission-line lens and
(b)~reference dielectric lens. A plane wave propagating in the
$+x$-direction illuminates the lenses. For clarity, the phase is
not plotted inside the lenses.} \label{phase}
\end{figure}

The lens with the given dimensions (shown in Fig.~\ref{lens})
having the previously described network structure and transition
layers was modelled with HFSS software. The simulation model is
shown in Fig.~\ref{phase} together with the HFSS model of a
reference dielectric lens having the same curvature and the
relative permittivity of $\varepsilon_r=4.66$. Three edges of the
simulation model are terminated with a perfectly matched layer
(PML) and in the symmetry plane (center of the lens) there is a
PMC boundary. First, the lenses shown in Fig.~\ref{phase} were
illuminated by a plane wave propagating in the $+x$-direction and
having electric field along the $z$-axis. The phase of the
simulated electric field in both systems is plotted in
Fig.~\ref{phase} for the frequency 2.4~GHz.

\begin{figure}[t!]
\centering {\epsfig{file=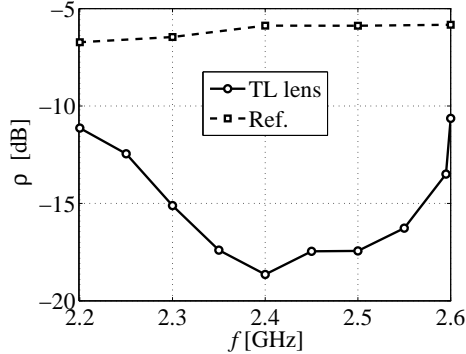,width=0.4\textwidth}}
\caption{Simulated reflectance of the lenses shown in
Fig.~\ref{phase}. The plane wave illuminating the lenses travels
in the $+x$-direction.}\label{rho_TL_diel}
\end{figure}

Clearly the proposed transmission-line lens refracts the incident
plane wave similarly as the homogeneous dielectric lens. The TL
lens reflectance, as compared to the reference lens, is lowered
significantly (by 4~dB or more) in a relative bandwidth of
approximately 16.7 percent, see Fig.~\ref{rho_TL_diel}. In order
to make a fair comparison between the two lenses shown in
Fig.~\ref{phase}, the losses inside the both lenses are equal,
i.e., the both dielectrics are lossless and the transmission-line
sections at the edges of the proposed TL lens (which are not
connected to the transition layer) are left open.

\begin{figure} [h!]
\centering {\epsfig{file=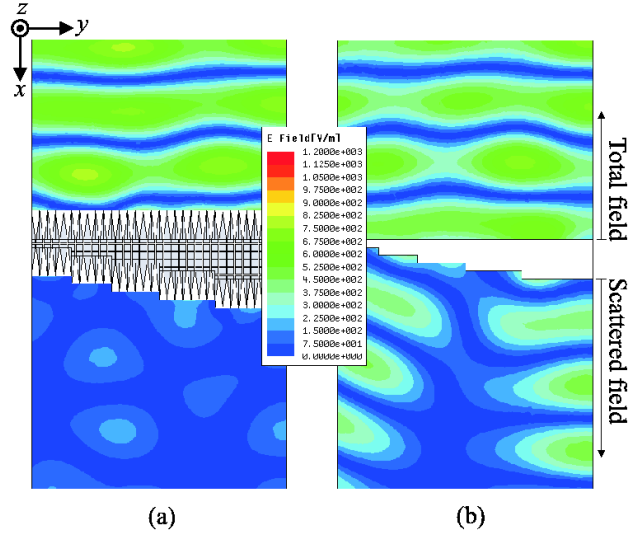, width=0.5\textwidth}}
\caption{Color Online. Simulated electric field at the frequency
2.4~GHz for (a) the transmission-line lens and (b) the reference
dielectric lens. A line source positioned at the focal point (see
Fig.~\ref{lens}) illuminates the lenses. On the source sides only
the scattered field is plotted in order to compare the magnitude
of the reflections from the lens interfaces. For clarity, the
fields are not plotted inside the lenses.} \label{Efield}
\end{figure}

To further demonstrate the focusing effect and the operation of
the designed TL lens, a line source was introduced at the expected
focal point. The expected focal point is located at the distance
$80d$ away from the lens as shown in Fig.~\ref{lens} (the reader
should note that the used simulation software allows sources to be
placed outside the finite simulation model). Snapshots of the
simulated electric field distributions are shown in
Fig.~\ref{Efield}, demonstrating that the cylindrical wavefronts
radiated by the source (with electric field along the $z$-axis)
are refracted in the correct way, creating a plane wave
(approximately) on the other side of the lens that travels to the
$-x$-direction. To illustrate the good impedance matching of the
TL lens, only the scattered fields are plotted on the source sides
of the lenses.

\section{Discussion and conclusions}

We have proposed a novel method for realizing a microwave lens
with highly reduced reflections as compared to traditional lenses
made of homogeneous dielectric materials. The lens is based on the
use of transmission-line networks that are matched to free space.
Since the refractive index of the network can be designed
separately from the network impedance, the network can be
impedance matched with free space while having high values of the
effective refractive index of the lens. We have presented simple
design equations for the dispersion and impedance of the
transmission-line networks, and have demonstrated the feasibility
of the proposed approach by simulations of a lens that focuses
incoming plane waves similar to a dielectric lens of the same size
and shape. In a certain frequency band the transmission-line lens
exhibits strongly mitigated reflection of power, as compared to
the reference dielectric lens.

\section*{Acknowledgements} This work has been partially supported
by the Academy of Finland and TEKES through the
Center-of-Excellence program. Pekka Alitalo wishes to thank the
Finnish Graduate School in Electronics, Telecommunications and
Automation (GETA), Tekniikan Edist\"amiss\"a\"ati\"o, the Finnish
Society of Electronics Engineers, and the Nokia Foundation for
financial support. Olli Luukkonen wishes to thank the Finnish
Society of Electronics Engineers, the Jenny and Antti Wihuri
Foundation, and the Nokia Foundation for financial support.

\clearpage


%
%
%


\newpage

\begin{thebibliography}{99}

\bibitem{Antenna_Theory}
R.E. Collin and F.J. Zucker, \textit{Antenna Theory Pt. 2}, New
York: McGraw-Hill, 1969.

\bibitem{Microwave_Optics}
S. Cornbleet, \textit{Microwave Optics}, New York: Academic Press,
1976.

\bibitem{Lo}
Y. T. Lo, S. W. Lee (eds.), {\it Antenna Handbook, Theory,
Applications and Design}, New York: Van Nostrand Reinhold Company,
1988.

\bibitem{Rotman}
W. Rotman, R. F. Turner, ``Wide-angle microwave lens for line
source applications,'' {\it IEEE Trans. Antennas Propag.},
pp.~623-632, Nov. 1963.

\bibitem{Patton}
W. T. Patton, ``Limited-scan arrays,'' {\it Proc. 1970 Phased
Array Antenna Symp.}, pp.~332-343, Dedham: Artech House, 1970.

\bibitem{Alitaloiwat08}
P. Alitalo, J. Vehmas, O. Luukkonen, L. Jylh\"a, and S.~Tretyakov,
``Microwave transmission-line lens matched with free space,'' to
appear in {\it Proc. 2008 IEEE International Workshop on Antenna
Technology: Small Antennas and Novel Metamaterials}, 2008.

\bibitem{Eleftheriades_book}
G.V. Eleftheriades and K.G. Balmain, \textit{Negative-Refraction
Metamaterials: Fundamental Principles and Applications}, Hoboken,
NJ: John Wiley \& Sons, 2005.

\bibitem{Caloz_book}
C. Caloz and T. Itoh, \textit{Electromagnetic Metamaterials:
Transmission Line Theory and Microwave Applications}, Hoboken, NJ:
John Wiley \& Sons, 2006.

\bibitem{AlitaloJAP}
P. Alitalo, S. Maslovski, and S.~Tretyakov, ``Three-dimensional
isotropic perfect lens based on $LC$-loaded transmission lines,''
\textit{J. Appl. Phys.}, vol.~99, p.~064912, 2006.

\bibitem{Alitalo_metmat}
P. Alitalo and S. Tretyakov, ``Subwavelength resolution with
three-dimensional isotropic transmission-line lenses,''
Metamaterials, vol.~1, no.~2, pp.~81-88, 2007.

\bibitem{AlitaloTAP}
P. Alitalo, O. Luukkonen, L. Jylh\"a, J. Venermo, and
S.~A.~Tretyakov, ``Transmission-line networks cloaking objects
from electromagnetic fields,'' arXiv:0706.4376, to appear in {\it
IEEE Trans. Antennas Propag.}








\end{thebibliography}
\end{document}